\documentclass[12pt]{article}
\usepackage{graphicx,amsmath,amssymb}

\newlength{\dinwidth}
\newlength{\dinmargin}
\def\lapproxeq{\lower .7ex\hbox{$\;\stackrel{\textstyle
<}{\sim}\;$}}
\def\gapproxeq{\lower .7ex\hbox{$\;\stackrel{\textstyle
>}{\sim}\;$}}
\def\gtrsim{\lower .7ex\hbox{$\;\stackrel{\textstyle
>}{\sim}\;$}}
\def\lesim{\lower .7ex\hbox{$\;\stackrel{\textstyle
<}{\sim}\;$}}

\def\be{\begin{equation}}
\def\ee{\end{equation}}
\def\bea{\begin{eqnarray}}
\def\eea{\end{eqnarray}}

\def\GeV{\rm GeV}
\def\J{J/\psi}

\begin{document}

\titlepage

\begin{flushright}


IPPP/07/50\\

DCPT/07/100\\

LTH 755\\

27 September 2007 \\

\end{flushright}

\vspace*{2cm}

\begin{center}

{\Large \bf Small $x$ gluon from exclusive $J/\psi$ production}

\vspace*{1cm} {\sc A.D. Martin}$^a$, {\sc C. Nockles}$^b$, {\sc
  M. Ryskin}$^{a, c}$ and {\sc T. Teubner}$^{b}$ \\

\vspace*{0.5cm}

$^a$ {\em Department of Physics and Institute for Particle Physics
  Phenomenology,\\ 

University of Durham, Durham DH1 3LE, U.K.}\\

$^b$ {\em Department of Mathematical Sciences,\\

 University of Liverpool, Liverpool L69 3BX, U.K.}

$^c$ {\em Petersburg Nuclear Physics Institute, Gatchina,
St.~Petersburg, 188300, Russia} \\

\end{center}

\vspace*{1cm}

\begin{abstract}
Exclusive $J/\psi$ production, $\gamma^* p \to J/\psi\,p$, offers a
unique opportunity to determine the gluon density of the proton in the
small $x$ domain.  We use the available HERA data to determine the
gluon distribution in the region $10^{-4} \lapproxeq x \lapproxeq
10^{-2}$ and $2 \lapproxeq Q^2 \lapproxeq 10~{\rm GeV}^2$, where the
  uncertainty on the gluon extracted from the global parton analyses
  is large.  The gluon density is found to be approximately flat at
  the lower scale; it is compared with those of recent global analyses.
\end{abstract}

\section{Introduction}
\begin{figure}
\begin{center}
\includegraphics[width=16cm]{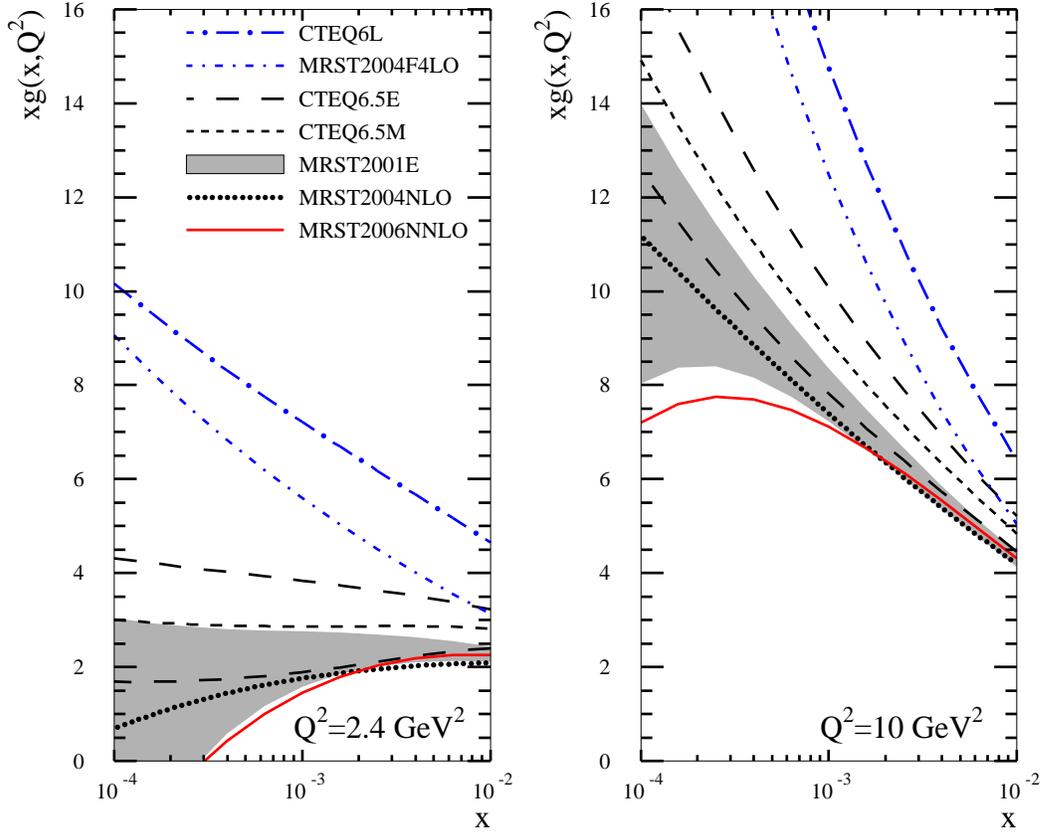}
\vspace{-8mm}
\caption{Comparison of recent global fits of the gluon distribution at
  small $x$ at leading (LO), next-to-leading (NLO) and next-to-next-to
  leading (NNLO) order, for the two scales $Q^2 = 2.4$ (left) and $10$
  GeV$^2$ (right panel).  LO gluons (dash dot) compared are
  CTEQ6L~\cite{pdfCTEQ6} and MRST2004F4LO~\cite{pdfMRST2004F}.  The two
  (long) dashed lines indicate the error estimate of the
  CTEQ6.5~\cite{pdfCTEQ6.5} gluon and the shaded band is the error
  band for the MRST2001~\cite{pdfMRST2001} global gluon.  Central values for
  the NLO global fits are from CTEQ6.5M (short dashed) and
  MRST2004NLO~\cite{pdfMRST2004} (dotted).  The solid line represents
  MRST2006NNLO~\cite{MRST07}.} 
\label{fig:globalgluons}
\end{center}
\end{figure}

Global analyses do not reliably determine the gluon for $x\lesssim
\mbox{\ a few\ } 10^{-2}$ at low, yet perturbative, $Q^2$ as shown in
Fig.~\ref{fig:globalgluons}. This is due partly to the lack of precise
structure function data for $x\lesssim 10^{-4}$ and mainly due to the
fact that the data included in global analyses actually probe the
quark distribution, while the gluon density is constrained by the
$\log Q^2$ dependence of the data, that is by the evolution. In the
low $x$ region the available $Q^2$ interval decreases and the accuracy
of the gluon determination becomes worse. 
The strong dependence of the global fits for the gluon on the order of
the analysis is clearly demonstrated in Fig.~\ref{fig:globalgluons}.
Note that the recent gluon from the MRST NNLO analysis~\cite{MRST07}
receives sizeable corrections both in size and shape compared to
the NLO fit, signalling a large uncertainty of the gluon in this
regime.  In this context it is also interesting to note that the gluon as
obtained from global fits can significantly change, both in
normalisation and shape, if small $x$ resummations are incorporated
into the analysis~\cite{White:2006yh}.

Data for the exclusive $\gamma^* p \to J/\psi\,p$ process offer an
attractive opportunity to determine the low $x$ gluon density in this
$Q^2$ domain, since here the gluon couples {\em directly} to the charm
quark and the cross section is proportional to the gluon density {\em
  squared}~\cite{Ryskin93}. Therefore the data are much more sensitive
to the behaviour of the gluon. The mass of the $c\bar c$ vector meson
introduces a relatively large scale, amenable to the perturbative QCD
(pQCD) description not only of large $Q^2$ diffractive
electroproduction, but also photoproduction of $J/\psi$. The available
$J/\psi$ data probe the gluon at a scale $\mu^2$ in the range $2 -
10$ GeV$^2$ for $x$ in the range $10^{-4} \lesssim x \lesssim
10^{-2}$; that is just the domain where other data do not constrain
the gluon reliably, see Fig.~\ref{fig:globalgluons}. It would be good
to have comparable data on exclusive $\Upsilon$ production to
determine the gluon at larger scales, but here the available data are
sparse, see Fig.~\ref{fig:upsilon} below. 

\section{Exclusive $J/\psi$ production at LO}
\begin{figure}
\begin{center}
\includegraphics[height=5cm]{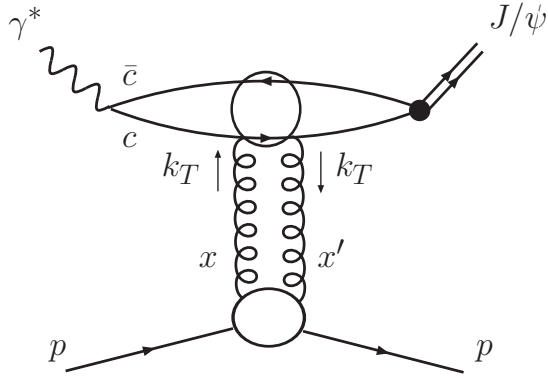}
\caption{Schematic picture of high energy elastic $\J$
  production, $\gamma^* p \to J/\psi\,p$. The factorised form
  follows since, in the proton rest frame, the formation time
  $\tau_f \simeq 2E_\gamma/(Q^2+M^2_{\J})$ is much greater than the
  $c\bar c$-proton interaction time $\tau_{\rm int}$.  In the case of
  the simple two gluon exchange shown here, $\tau_{\rm int} \simeq R$,
  where $R$ is the radius of the proton.} 
\label{fig:feyndiagram}
\end{center}
\end{figure}

To lowest order the $\gamma^*p \to \J \, p$ amplitude can be factored
into the product of the $\gamma \to c\bar c$ transition, the
scattering of the $c\bar c$ system on the proton via (colourless)
two-gluon exchange, and finally the formation of the $\J$ from the
outgoing $c\bar c$ pair. The crucial observation is that at high
$\gamma p$ centre-of-mass energy, $W$, the scattering on the proton
occurs over a much shorter timescale than the $\gamma \to c\bar c$
fluctuation or the $\J$ formation times, see
Fig.~\ref{fig:feyndiagram}.  Moreover, at leading logarithmic accuracy,
this two-gluon exchange amplitude can be shown to be directly
proportional to the gluon density $xg(x,\bar Q^2)$ with
\begin{equation}
{\bar Q^2}~=~(Q^2+M^2_{\J})/4, ~~~~~~~~~x~=~(Q^2+M^2_{\J})/(W^2+M^2_{\J}).
\end{equation}
$Q^2$ is the virtuality of the photon and $M_{\J}$ is the rest mass of
the $\J$.  To be explicit, the lowest-order formula is~\cite{Ryskin93} 
\begin{equation}
\frac{{\rm d}\sigma}{{\rm d}t}\left( \gamma^* p \to J/\psi ~p \right)
     {\Big |}_{t=0} = \frac{\Gamma_{ee}M^3_{\J}\pi^3}{48\alpha}\,
     \left[\frac{{\alpha_s(\bar Q^2)}}{\bar Q^4}xg(x,\bar
     Q^2)\right]^2 \left(1+\frac{Q^2}{M^2_{\J}}\right),
\label{eq:lo}
\end{equation}
where $\Gamma_{ee}$ is the electronic width
of the $J/\psi$.

In the leading logarithmic approximation, the integral over the transverse
momentum $k_T$ of the $t$-channel gluons, see
Fig.~\ref{fig:feyndiagram}, gives rise to the integrated gluon density
$g(x,\bar Q^2)$. As usual in collinear factorisation, the $k_T$
dependence of the integral is completely absorbed in the input gluon
distribution (of the global analyses), taken at the factorisation
scale $\bar Q^2$. The integral over the charm quark loop is expressed
in terms of the electronic width, $\Gamma_{ee}$, of $J/\psi$, and the
$Q^2/M_{J/\psi}^2$ term in the final brackets reflects the
contribution of the longitudinally polarised incoming
$\gamma^*$. Equation~(\ref{eq:lo}) gives the differential cross
section at zero momentum transfer, $t=0$.  To describe data integrated
over $t$, the integration is carried out assuming $\sigma \sim \exp(-bt)$
with $b$ the experimentally measured slope parameter.  Throughout this
work the value 
\begin{equation}
b = 4.5 {\rm\ GeV}^{-2}
\label{eq:slope}
\end{equation}
is used, which is in agreement with~\cite{H1,ZEUS1,ZEUS2}.\footnote{We
  neglect a slight energy dependence of $b$ which is only observed for
  photoproduction, but which is of the order of differences between
  measurements of 
  the two experiments H1 and ZEUS.  The possible uncertainty is
  smaller than, or at most comparable to, other approximations used.}
Thus it becomes possible to extract the gluon density $g(x,\bar Q^2)$
directly from the measured diffractive $J/\psi$ cross section. 

\section{Corrections to LO exclusive $J/\psi$ production}
\label{sec:nlo}
Expression (\ref{eq:lo}) is a simple, first approximation, justified
in the LO collinear approximation using the non-relativistic $J/\psi$
wave function. The relativistic corrections were intensively discussed
in~\cite{RRML,fks}. The problem is that, simultaneously with the
relativistic description of the $c$ quarks, one needs to consider
higher order Fock component $c\bar c g$ states of
$J/\psi$. Hoodbhoy~\cite{Hood} has studied these two effects to order
$v^2/c^2$.  He has shown that relativistic corrections to
(\ref{eq:lo}), written in terms of the experimentally measured
$\Gamma_{ee}$, are small, $\sim{\cal O}(4\%)$, see~\cite{Hood}. So we
do not account further for relativistic corrections below.

NLO corrections arise, first, from an explicit integration over the
  gluon $k_T$, which goes beyond the leading log contribution, arising
  from ${\rm d}k_T^2/k_T^2$, and, second, from more complicated
diagrams with one additional loop. To perform the explicit $k_T$ integration
we have to use the unintegrated gluon distribution, $f(x,k_T^2)$,
which is related to the integrated gluon by
\begin{equation}
x g(x,\mu^2) = \int_{Q_0^2}^{\mu^2} \frac{{\rm
    d}k_T^2}{k_T^2}\,f(x,k_T^2) + c(Q_0^2)\,.
\label{eq:unintgluon}
\end{equation}
Of course, the infrared contribution cannot be treated perturbatively,
and so we have introduced a lower limit $Q_0^2$ for the $k_T^2$
integral in the $J/\psi$ production amplitude, 
\begin{equation}
\left[\frac{{\alpha_s(\bar Q^2)}}{\bar Q^4}xg(x,\bar Q^2)\right]
\:\longrightarrow\: 
\int_{Q_0^2}^{(W^2-M_{\J}^2)/4} 
\frac{{\rm d}k_T^2\,\alpha_s(k_T^2)}{\bar Q^2 (\bar Q^2 + k_T^2)} \, 
\frac{\partial \left[ xg(x,k_T^2)\right]}{\partial k_T^2} \: + \: 
  \frac{\alpha_s(Q_0^2)}{\bar Q^4} xg(x,Q_0^2)\,.
\label{eq:nlointegral}
\end{equation}
Expression (\ref{eq:nlointegral}) replaces the factor $\alpha_s(\bar
Q^2)xg(x,\bar Q^2)/\bar Q^4$ in the LO result (\ref{eq:lo}). To be
precise, the unintegrated distribution $f$ embodies the Sudakov factor
$T(k_T^2,\mu^2)$ such that~\cite{mrt2}
\begin{equation}
f(x,k_T^2) = \partial\left[ xg(x,k_T^2) T(k_T^2,\mu^2)
  \right]/\partial\ln k_T^2\,. 
\label{eq:deftfactor}
\end{equation}
Thus $c(Q_0^2)$ in Eq.~(\ref{eq:unintgluon}) is given by $xg(x,Q_0^2)
T(Q_0^2,\mu^2)$ and correspondingly for (\ref{eq:nlointegral}),
$xg(x,Q_0^2)\to xg(x,Q_0^2) T(Q_0^2,\mu^2)$.  In our numerics we have
chosen $\mu^2 = \bar Q^2$.  However, in the amplitude
(\ref{eq:nlointegral}), the dominant contribution comes from the
region of $k_T \sim \bar Q$ where $T(k_T^2,\mu^2)$ is close to
unity.  The inclusion of the $T$ factor may be considered as an ${\cal
  O}(\alpha_s)$ correction to the gluon density and suppresses the
gluon in our analysis by  1.7\% for photoproduction at $x=10^{-3}$.
The contribution coming from $k_T < Q_0$ is written in terms of the
integrated gluon $g(x,Q_0^2)$, that is the infrared part is absorbed
into the input distribution at the `transition' scale $Q_0$. 

Of course, at low $Q^2$ the gluon extracted from a global analysis may
be affected by the presence of absorptive corrections which are
usually neglected.  Here, the absorptive corrections are expected to
be smaller.  The transverse size, $r$, of the $q\bar q$ dipole
produced by the `heavy' photon in DIS has a logarithmic distribution
$\int {\rm d}r^2/r^2$ starting from $1/Q^2$ up to some hadronic
scale.  In the case of $\J$ production the size of the $c\bar c$
dipole is limited by the size of the $\J$ meson.  Even in
photoproduction it is of the order of $1/\bar Q^2$.  Since the
probability of rescattering is proportional to $r^2$, we anticipate a
much smaller absorptive effect.

A more detailed analysis of the NLO
corrections was done in~\cite{ikp,issk}. Part of these
corrections generates the running of $\alpha_s$, while part is similar
to gluon Reggeization in the BFKL approach. Indeed, for $\J$
electroproduction it was shown~\cite{issk} using the conventional
collinear factorisation scheme, that there is a NLO correction of the form
\be
\frac{3\alpha_s}{\pi}~\ln\left(\frac{1}{x}\right)\,\ln\left(\frac{\bar
  Q^2}{\mu^2}\right)\,. 
\label{eq:DLog}
\ee
In the $k_T$ factorisation approach such a correction may be included
by replacing the $t$-channel gluon by the Reggeized gluon. However
this correction vanishes with a natural choice of the factorisation
scale, $\mu^2 = \bar Q^2$, which was adopted in our
prescription~\cite{lmrt,mrt1,mrt2}.  One therefore has reason to believe
that the $k_T$ factorisation approach accounts for a major part of the
NLO effect, and that the resulting `NLO' gluon may be compared to that
in a set of NLO global partons.\footnote{The global partons are
  defined in the $\overline{\rm MS}$ regularization scheme. Our
  partons should also be considered to be in the $\overline{\rm MS}$
  scheme, since we use the $\overline{\rm MS}$ definition of
  $\alpha_s$, and moreover the factorisation scale which provides the
  cancellation of the $\alpha_s \ln 1/x$ correction is also specified
  in the $\overline{\rm MS}$ scheme.}  Therefore we shall refer to the
resulting distributions as NLO gluons. 

One also needs to account for the fact that the two gluons exchanged
carry different fractions $x, x^{\prime}$ of the light-cone proton
momentum, see Fig.~\ref{fig:feyndiagram}. That is one has to use the
generalised (skewed) gluon distribution.\footnote{In the formal
  analysis of the NLO contributions, there are effects arising from
  integrated quarks which generate gluons which then couple to the
  charm quark. In terms of our unintegrated gluon ($f$) description
  this should be considered as a NLO correction to the evolution of
  $f$. Since we do not consider the evolution, but just parametrise
  the scale dependence of the gluons (see below), this correction is
  outside our analysis.} 
In our case $x^{\prime} \ll x \ll 1$, and the skewing effect can be
well estimated from~\cite{shuvaev} 
\begin{equation}
R_g = \frac{2^{2\lambda + 3}}{\sqrt{\pi}}
\frac{\Gamma(\lambda + \frac{5}{2})}{\Gamma(\lambda + 4)}
\label{eq:defrg}
\end{equation}
where $\lambda(Q^2) = \partial\left[\ln(xg)\right]/\partial\ln
x$. That is in the small $x$ region of interest we take the gluon to
have the form $xg \sim x^{-\lambda}$. 

Recall that the integral (\ref{eq:nlointegral}) was written for the
discontinuity (i.e. for the imaginary part) of the amplitude shown in
Fig.~\ref{fig:feyndiagram}. The real part may be determined using a
dispersion relation. In the low $x$ region, for our positive-signature
amplitude $A \propto x^{-\lambda} + (-x)^{-\lambda}$, the dispersion
relation can be written in the form 
\begin{equation}
\frac{{\rm Re} A}{{\rm Im} A} \simeq \frac{\pi}{2}\lambda \simeq
\frac{\pi}{2} \frac{\partial\ln A}{\partial\ln(1/x)} \simeq
\frac{\pi}{2} \frac{\partial\ln\left(xg(x,\bar
  Q^2)\right)}{\partial\ln(1/x)}\,. 
\label{eq:defre}
\end{equation}

Both corrections lead to an enhancement of the cross section.

\section{Determination of the gluon from $J/\psi$ data}
In the following, we present fits to the data for exclusive $\J$
production from HERA using the perturbative description discussed
above. 

In the low $x$ region it is expected that the $x$ dependence of the
gluon density $x g(x,Q^2)$ is well approximated by the form
$x^{-\lambda}$.  However, the evolution in $Q^2$ modifies this
behaviour, enlarging the power $\lambda$ as $Q^2$ increases.  In
particular, in the double leading log (DLL) approximation, we have the
asymptotic form 
\be
x g
\sim \exp\left(\sqrt{\frac{4\alpha_s N_c}{\pi} \ln(1/x)\,\ln
  Q^2}\right).
\ee
 Thus we need a $Q^2$ dependent parametrisation. However, in the
 limited region of $Q^2$ covered by the exclusive $J/\psi$ data, it is
 sufficient to use a simple parametric form\footnote{Such a form has
   already successfully been used in~\cite{MRW} for the analysis of
   inclusive diffractive DIS data.} 
\be
xg(x,\mu^2)\ =\ N x^{-\lambda} \qquad {\rm with}\ \ \lambda = a +
b\,\ln(\mu^2/0.45 {\rm\ GeV}^2)\,. 
\label{eq:gluonansatz}
\ee
The free parameters $N$, $a$ and $b$ are determined by a non-linear
$\chi^2$ fit to the exclusive $J/\psi$ data from H1~\cite{H1} and
ZEUS~\cite{ZEUS2,ZEUS3}.\footnote{We also performed fits for the data
  from the H1 and ZEUS collaborations separately.  These fits
  typically have a smaller $\chi_{\rm min}^2$, signalling a slight
  incompatibility between the data.  However, they lead to similar
  results for the gluon.  As the combined fit is very satisfactory
  ($\chi^2_{\rm min}/{\rm d.o.f. < 1}$), we will not discuss fits of
  individual data sets in the following.} 
This three-parameter form provides enough flexibility to accurately
describe the $x$ and $Q^2$ behaviour of $\J$ production in the limited
domain covered by the $J/\psi$ data, namely $10^{-4}<x<10^{-2}$ and
$2<Q^2<8~\GeV^2$, so we will use exactly Eq.~(\ref{eq:gluonansatz})
for the LO fit. 

However, for the NLO approach, where we have the $k_T^2$ integral
(\ref{eq:nlointegral}) which runs up to the kinematical limit $k_T^2 =
(W^2 - M_{\J}^2)/4$, we face the problem of a badly convergent $k_T^2$
integral.  Indeed the low $x$ gluons $xg \sim x^{-\lambda}$ obtain a
large anomalous dimension $\gamma$.  Since $x^{-b\ln k_T^2} =
(k_T^2)^{b\ln(1/x)}$ we get $\gamma = b\ln(1/x)$.  On the other hand,
the expression (\ref{eq:nlointegral}) was calculated with running
$\alpha_s$.  Recall that the scale dependence of the power $\lambda$
was generated by the evolution of the form $\lambda = \int
\alpha_s(q^2) \frac{{\rm d}q^2}{q^2}$.  Accounting for the running
$\alpha_s$, it is natural to replace the second term in the parametric
form (\ref{eq:gluonansatz}) by $\ln\ln(\mu^2/\Lambda_{\rm QCD}^2)$.
Therefore, for the NLO fit, we use
\be
xg(x,\mu^2)\ =\ N x^{-\lambda} \qquad {\rm with}\ \ \lambda = a +
b\,\ln\ln(\mu^2/\Lambda_{\rm QCD}^2)\,. 
\label{eq:gluonansatznlo}
\ee

\begin{figure}
\begin{center}
\vspace{-6mm}
\includegraphics[width=17cm]{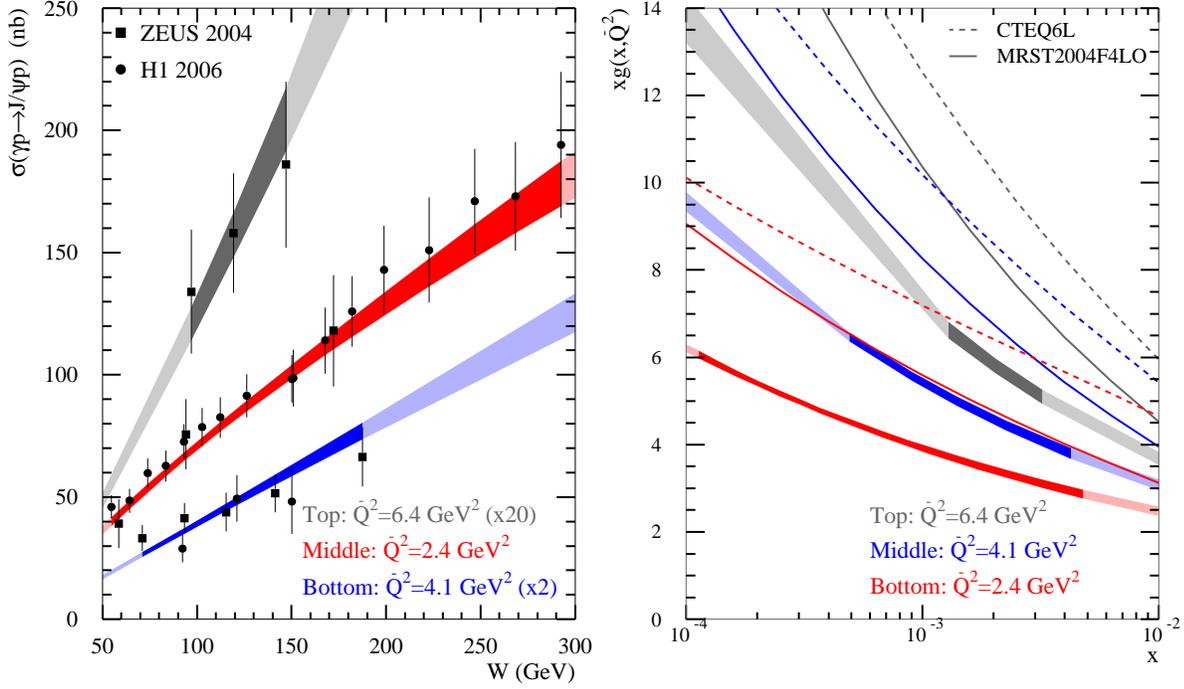}
\vspace{-10mm}
\caption{Leading order fit of elastic $\J$ data as described in the
  text.  Left panel: cross section compared to some of the
  H1~\cite{H1} and ZEUS~\cite{ZEUS2,ZEUS3} data, with values for $\bar
  Q^2$ as indicated; right panel: gluon compared to global fits for
  scales as indicated, where the solid (dashed)
  lines are the MRST2004F4LO~\cite{pdfMRST2004F}
  (CTEQ6L~\cite{pdfCTEQ6}) results.  The width of the bands displays
  the uncertainty of the cross section and fitted gluon respectively,
  whereas the darker shaded areas indicate the region of the available
  data.} 
\label{fig:lo}
\end{center}
\end{figure}
\begin{figure}
\begin{center}
\vspace{-6mm}
\includegraphics[width=17cm]{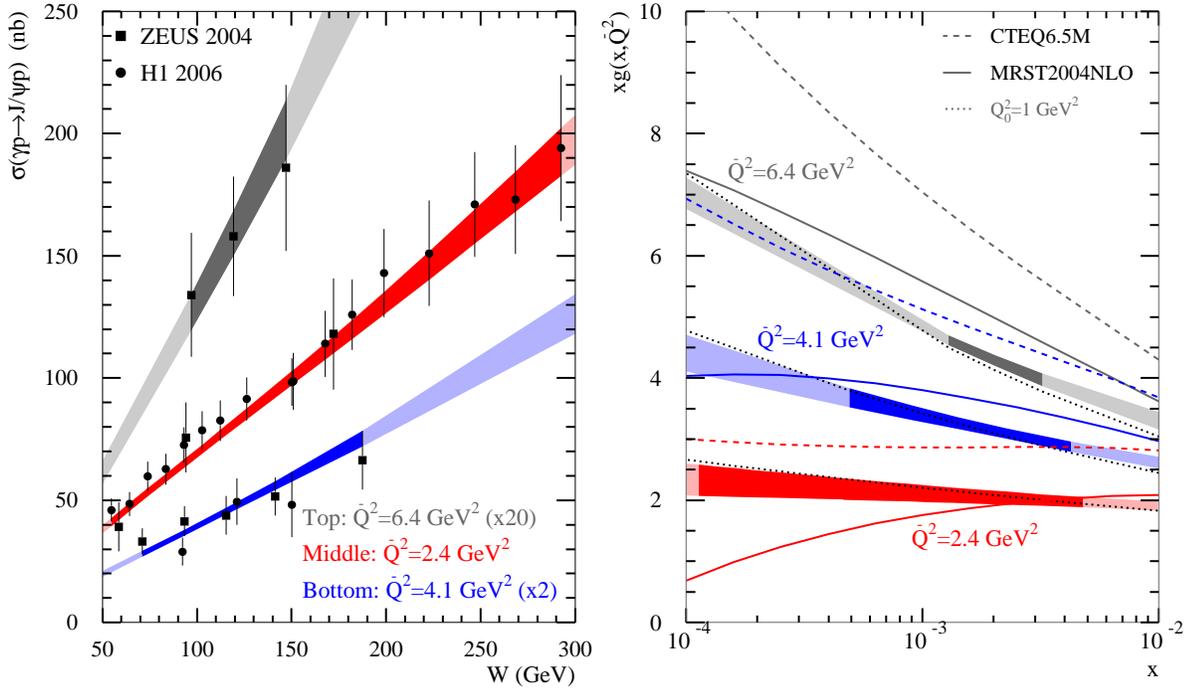}
\vspace{-10mm}
\caption{As Fig.~\ref{fig:lo}, but for the next-to-leading order fit
  and comparing to NLO global fits from CTEQ6.5M~\cite{pdfCTEQ6.5} and
  MRST2004~\cite{pdfMRST2004}.} 
\label{fig:nlo}
\end{center}
\end{figure}

The gluon densities obtained from the LO fit, using (\ref{eq:lo}) with
(\ref{eq:gluonansatz}), together with the skewing and real part
corrections, to the exclusive $J/\psi$ data~\cite{H1,ZEUS2,ZEUS3} are
shown in Fig.~\ref{fig:lo}.\footnote{Note that in
  Figs.~\ref{fig:lo},~\ref{fig:nlo} only a subset of 51 data
  points used in the fits is displayed, and there are data points at
  up to $\langle Q^2 \rangle = 22.4$ GeV$^2$, corresponding to $\bar
  Q^2 = 8$ GeV$^2$.} 
Inclusion of these corrections gives a
22\% suppression of the gluon for photoproduction at $x~=~10^{-3}$,
with the skewing correction giving the dominant suppression of 18\%.
We use a 1-loop running $\alpha_s$ with $\alpha_s(M_Z^2)=0.118$. 
In the analysis, error bands on the gluon and cross section are
generated using the full covariance matrix for the fitted parameters,
where as input we have added the statistical and systematic 
experimental errors of the data in quadrature.
Compared to the gluons from the global fits, the gluon from our LO fit
is similar in shape, but slightly smaller in normalisation and shows
less rise towards smaller $x$ with growing scales. 

We now present the results of our NLO fit, which we obtain by
modifying Eq.~(\ref{eq:lo}) by help of (\ref{eq:nlointegral}) with
2-loop running $\alpha_s$, and replacing the parametrisation
(\ref{eq:gluonansatz}) by (\ref{eq:gluonansatznlo}).  Of course we
also include skewing and real part corrections as before, and the $T$ 
factor as discussed above. 
\begin{table}
\begin{center}
\begin{tabular}{c|c c c c}      
\hline\hline 
     & $N$ & $a$ & $b$ & $\chi^2_{\rm min}/{\rm d.o.f.}$\\ \hline
  LO & $ 0.99 \pm 0.09 $ & $ 0.051 \pm 0.012$ & $ 0.088 \pm 0.005$ & 0.9\\
 NLO & $ 1.55 \pm 0.18 $ & $ -0.50 \pm 0.06 $ & $ 0.46 \pm 0.03 $ & 0.8\\
\hline\hline 
\end{tabular}
\end{center}
\vspace{-2mm}
\caption{Values of the three parameters of the LO and NLO gluon fits and
  corresponding $\chi^2_{\rm min}/{\rm d.o.f.}$}
\label{parameters}
\end{table}
\begin{table}
\begin{center}
\begin{tabular}{c||c|c|c}
\hline \hline
$Q^2 (\GeV^2)$ & $\lambda_{\J}$ & $\lambda_{\rm MRST}$ &
$\lambda_{\rm CTEQ}$\\ 
\hline
$2.4$ & $0.04$ & $-0.17$ $\{-1.07,\,-0.16,\,0.00\}$ & $0.01$
$\{0.04,\,0.00,\,0.05\}$ \\ 
$4.1$ & $0.11$ & $ 0.06$ $\{-0.03,\,0.07,\,0.16\}$ & $0.13$
$\{0.14,\,0.13,\,0.19\}$ \\ 
$6.4$ & $0.16$ & $ 0.15$ $\{0.09,\,0.15,\,0.24\}$  & $0.19$
$\{0.18,\,0.19,\,0.27\}$ \\ 
$8.0$ & $0.19$ & $ 0.18$ $\{0.13,\,0.18,\,0.27\}$  & $0.21$
$\{0.20,\,0.22,\,0.30\}$ \\ 
\hline \hline
\end{tabular}
\end{center}
\vspace{-2mm}
\caption{The values of the power of the gluon,
  $\lambda$, at four $Q^2$ values, for our NLO fit to elastic
  $\J$ production data compared to two global
  fits~\cite{pdfCTEQ6.5,pdfMRST2004}.  The numbers for MRST and CTEQ
  are obtained through a fit in the range $x = 10^{-4} \ldots 10^{-2}$,
  assuming $xg \sim x^{-\lambda}$ with an $x$ independent $\lambda$, whereas
  the values in curly brackets are the logarithmic derivatives of the gluons
  at $x = \{ 10^{-4},\,10^{-3},\,10^{-2}\}$, respectively.} 
\label{lambda}
\end{table}
Figure~\ref{fig:nlo} shows our fit using $Q_0^2=2 {\rm\ GeV}^2$ and
$\Lambda_{\rm QCD}^2 = 0.09 {\rm\ GeV}^2$.  The dotted lines represent
the central values of the gluons obtained using $Q_0^2=1 {\rm\
  GeV}^2$.  The NLO gluon fit shows a better matching to the global
gluons at $x~=~10^{-2}$ than the LO gluon obtained.  The analysis of
the exclusive $\J$ data indicates that, at the larger scales, the small $x$
behaviour of the gluon distribution is slightly flatter than that of
the global analyses, both in their $x$ behaviour and in their scale
dependence of $\lambda(Q^2)$.  At low scales, the gluon obtained from
the $\J$ data still rises with decreasing $x$, especially in contrast
to the MRST fit.  For completeness we also present the values of the
parameters for the LO and NLO fits (Tab.~\ref{parameters}).  As can be
seen from Figs.~\ref{fig:lo}, \ref{fig:nlo} and the $\chi^2_{{\rm
    min}}/{\rm d.o.f.}$ values quoted in Tab.~\ref{parameters}, our simple
ansatz for the form of the gluon, $xg \sim x^{-\lambda}$, using an $x$
independent power $\lambda$, works very well.  To
quantify the $x$ and $Q^2$ behaviour, we tabulate the values of the
power of the gluon, $\lambda$, from
(\ref{eq:gluonansatznlo}) at four $Q^2$ values, compared to values
estimated from MRST2004NLO~\cite{pdfMRST2004} and CTEQ6.5M
gluons~\cite{pdfCTEQ6.5} (Tab.~\ref{lambda}).  As is evident from
Fig.~\ref{fig:nlo} and Tab.~\ref{lambda}, our NLO gluon fit seems to
be incompatible with the strength of evolution of the MRST global fit
(e.g. at $x=10^{-3}$ our gluon increases by a factor $2.7$ from $Q^2 =
2.4$ to $8$ GeV$^2$ compared to $3.7$ for the MRST2004NLO gluon);
however, there is fair agreement in the evolution between our gluon
and the CTEQ fit (CTEQ evolves by a factor $2.8$ in the same regime,
although in absolute normalisation our NLO $J/\psi$ prediction for the
gluon agrees, on average, much better with the MRST NLO prediction).
Of course, this should be seen in light of the large
uncertainties of both the MRST and CTEQ gluons at small $x$ and
scales, see Figs.~\ref{fig:globalgluons} and \ref{fig:nlo}.  This
uncertainty persists at
the largest scales probed in our fit.  As these scales are rather low,
the accuracy of the DGLAP approach may already be seriously affected
by small $x$ effects and absorptive and power suppressed corrections. 

Of course, in the $k_T$ factorisation approach there is some 
uncertainty arising from the infrared cut-off $Q_0$, below which we
cannot consider the $k_T$ integration literally. We have to express this
low $k_T$ contribution in terms of the gluon integrated over $k^2_T$
up to $Q_0^2$. However, with our prescription for unintegrated partons,
these two contributions match smoothly to each other. The ambiguity
due to the choice of $Q_0$ is quite small, as illustrated by the
closeness of the dotted lines to our NLO gluons in
Fig.~\ref{fig:nlo}. 

Note that the difference between the LO and NLO gluons is large at the
smallest $x$ values, both in the global parton analyses and for the
gluons obtained from elastic $\J$ production. For the global
analyses this is due to the absence of the photon-gluon coefficient
function at LO. At LO, the photon couples only to the quark parton,
which is produced from a gluon at some scale $q^2 \ll Q^2$, due to the
strong ordering in $k_T$. At low $x$ the gluon grows with $q^2$, and,
to provide the measured values of the proton structure function $F_2$,
we need a much larger gluon distribution in the LO formalism. At NLO
the photon-gluon coefficient function is present, which provides a
direct photon-gluon parton coupling at the scale $Q^2$, and, more
importantly, a $1/x$ divergence appears in the quark-gluon splitting
function, which accelerates the quark evolution and in turn requires
less gluon. 
In elastic $\J$ production we have an analogous situation. By carrying
out the $k_T$ integration, we include the interaction with the gluon
at large scales of the order of $Q^2+M^2_{\J}$. Moreover, part of this
integral has $k_T^2 \gg Q^2$, and may be regarded as one step of
backward evolution. 
In summary, in DGLAP analyses based on collinear factorisation,
the large change in the gluon distribution in going from LO to NLO is
due to the strong ordering in $k_T$ and the absence of an additional
loop integral in the LO coefficient function and parton evolution.
Inclusion of higher-order terms beyond NLO is expected to give a much
smaller effect.  They will mainly affect the normalisation and not the
$x$ dependence of the gluon. 

The higher-order corrections in our approach are ${\cal O}(\alpha_s)$, but
may be enhanced by large logarithms.  The corrections which are
enhanced by $\ln(1/x)$, and which may lead to some $x$ dependence,
are absorbed in the form of the gluon distribution by choosing the
appropriate factorisation scale $\mu^2 = \bar Q^2$, see
Eq.~(\ref{eq:DLog}).  Then, 
in our NLO approach, the part of the $k_T$ integral which may be
logarithmically large, is accounted for explicitly in the $k_T$
factorisation formalism which is used to obtain the $k_T$ integral.
After this, those higher-order corrections, which are not included in
the $k_T$ integral, are concentrated in the domain $k_T \sim \mu =
\bar Q$.  Thus the scale dependence of these higher-order corrections
is driven mainly by the scale dependence of the running of $\alpha_s$
which we take into account.  The fact that the gluons obtained from
our analysis turn out to be close to the gluon distributions
coming from the global analyses for $x \simeq 10^{-2}$, where they are
fairly stable, indicates that the omitted higher-order corrections are
indeed small. 

\begin{figure}
\begin{center}
\vspace{-6mm}
\includegraphics[width=12cm]{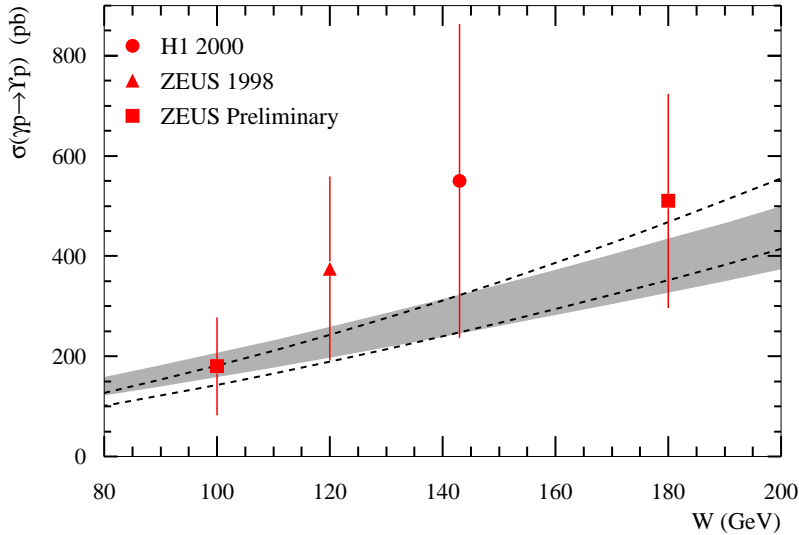}
\vspace{-6mm}
\caption{Prediction of elastic $\Upsilon$ photoproduction, using our
  LO and NLO gluon, compared to data.  The dotted lines indicate the
  error band of the LO prediction, whereas the shaded band is our NLO
  prediction.  The data points are the published
  ZEUS~\cite{ZEUSupsilon} and H1~\cite{H1upsilon} results, and the
  preliminary results from ZEUS~\cite{ZEUSupsprel07}.} 
\label{fig:upsilon}
\end{center}
\end{figure}
We have extended our framework to predict elastic $\Upsilon$
photoproduction, using our LO and NLO gluon with our cross section
formulae including corrections and $T$ factor, see
Fig.~\ref{fig:upsilon}.  Although the data is sparse, the cross
section predictions are reasonable. 

We conclude that this new information coming from our NLO analysis of
elastic $\J$ production data, in which the $c\bar{c}$ couple {\it
  directly} to the low $x$ gluon parton and where the cross section is
proportional to the {\it square} of the gluon, is especially valuable
to constrain the small $x$ behaviour of the gluon distribution.  The
accuracy of the elastic $\J$ data is now sufficient, as indicated by
the error bands (arising from the experimental uncertainties) on the
extracted gluon distribution shown in
Fig.~\ref{fig:nlo}, to improve our knowledge of the gluon distribution
at small $x$, $x \gapproxeq 10^{-4}$, considerably.  This is in
comparison to the small $x$ behaviour of the `global' gluon
distributions, which is not well constrained by the inclusive DIS
structure function data.\\ 

\noindent
{\bf\large Acknowledgements}

\noindent
We would like to thank Robert Thorne for valuable discussions.  TT
thanks the UK Science and Technology Facilities Council for an
Advanced Fellowship.  The work of MR was supported in part by the
Federal Program of the Russian Ministry of Industry, Science and
Technology, RSGSS-5788.2006.02.

\end{document}